\documentclass[preprint,onecolumn,nofootinbib]{revtex4}
\usepackage[colorlinks=true,linkcolor=blue,urlcolor=blue,filecolor=black,citecolor=red,pdfstartview=FitV,pdftitle={},pdfsubject={},pdfkeywords={},pdfpagemode=None,bookmarksopen=true]{hyperref}
\usepackage{graphicx}
\usepackage{amsmath}
\usepackage{amsfonts}
\usepackage{amssymb,ulem}
\usepackage{color,xcolor}%
\usepackage{CJK}
\usepackage{subfigure}
\usepackage{amsthm,amsmath,amssymb}
\usepackage{mathrsfs}
\usepackage{multirow}

\usepackage{dcolumn}
\usepackage{float}
\begin{document}
\title{Quasinormal modes of a regular black hole with sub-Planckian curvature}
\author{Dan Zhang $^{1}$}
\thanks{danzhanglnk@163.com}
\author{Huajie Gong$^{2}$}
\thanks{huajiegong@qq.com}
\author{Guoyang Fu$^{3}$}
\thanks{FuguoyangEDU@163.com}
\author{Jian-Pin Wu$^{2}$}
\thanks{jianpinwu@yzu.edu.cn, corresponding author}
\author{Qiyuan Pan$^{1}$}
\thanks{panqiyuan@hunnu.edu.cn, corresponding author}
\affiliation{$^1$~ Department of Physics, Key Laboratory of Low Dimensional Quantum Structures and Quantum Control of Ministry of Education, Institute of Interdisciplinary Studies, and Synergetic Innovation Center for Quantum Effects and Applications, Hunan Normal University,  Changsha, Hunan 410081, China
\\
$^2$~Center for Gravitation and Cosmology, College of Physical Science and Technology, Yangzhou University, Yangzhou 225009, China
\\
$^3$ Department of Physics and Astronomy, Shanghai Jiao Tong University, Shanghai 200240, China}
	
\begin{abstract}
		
This paper explores the properties of the quasinormal modes (QNMs) of a regular black hole (BH) characterized by a Minkowski core and sub-Planckian curvature. When focusing on a special case, this regular BH exhibits identical large-scale behavior with the Hayward BH and some loop quantum gravity corrected (LQG-corrected) BH. A notable characteristic of the QNMs in this regular BH is the pronounced outburst of overtones when compared to the Schwarzschild BH (SS-BH). This outburst can be attributed to the deviation from the SS-BH in the near-horizon geometry region due to the quantum gravity effect. Furthermore, we compare the QNM properties of the regular BH with those of the Hayward BH and the LQG-corrected BH. A similar phenomenon of overtone outburst is observed in the modes of the overtone. As a conclusion, the QNMs may be a powerful tool for detecting the quantum gravity effect and distinguishing different BH models.

\end{abstract}
	
\maketitle
	
\section{Introduction}\label{sec-intro}
	
General relativity (GR) is plagued by the well-known singularity issue, wherein the scalar curvature becomes infinitely large at the core of the black hole (BH) \cite{Hawking:1966sx,Penrose:1964wq,Joshi:2011rlc,Goswami:2005fu,Janis:1968zz}. Singularities signify GR's failure. Therefore, it makes sense to anticipate that a quantum theory of gravity would eventually take hold in these regions and solve the singularity issue. Prior to the establishment of a full theory of quantum gravity, physicists intensively investigated various non-singular BHs at the phenomenological level, commonly known as regular BHs. These BHs can be constructed by incorporating quantum gravity effects or exotic matter that typically violates the standard energy conditions. Typically, these regular BHs usually do not fulfill the vacuum Einstein equations. For comprehensive reviews on regular BHs, we can refer to  \cite{Bambi:2023try,Vagnozzi:2022moj,Lan:2023cvz}.
	
In terms of the asymptotic behavior near the center of the regular BHs, we can classify them into two types: one featuring a de-Sitter (dS) core, and the other featuring a Minkowski core. The regular BH with dS core includes the well-known Bardeen BH \cite{Bardeen:1968}, Hayward BH \cite{Hayward:2005gi}\footnote{We would like to mention that in \cite{Boos:2023icv}, the authors propose a novel regular BH solution, expanding upon the Hayward class significantly. This novel solution incorporates mass-dependent regulators, enabling notable, percent-level effects in observables for regular astrophysical BHs.} and Frolov BH \cite{Frolov:2016pav}.  The key characteristic of these regular BHs is that their Kretschmann scalar curvature remains finite everywhere.
	
The BH with Minkowski core was initially postulated in \cite{Xiang:2013sza}, featuring a Newton potential that is exponentially suppressed. Subsequently, this solution has been extended to a class of regular BHs with various forms of exponential potentials in \cite{Culetu:2013fsa,Culetu:2014lca,Rodrigues:2015ayd,Simpson:2019mud,Ghosh:2014pba,Ghosh:2018bxg,Li:2016yfd,Martinis:2010zk}. Additionally, as the Hawking temperature vanishes during the final phase of evaporation for this type of regular BH, it provides a more realistic depiction of the final phase of BH evaporation and a more reasonable behavior of its remnant. However, it is important to note that these regular BHs featuring a Minkowski core are meaningful only for BHs with relatively small mass at the Planck scale or during the last phase of BH evaporation.
	
Furthermore, based on the generalized uncertainty principle over curved spacetime, a novel regular BH with Minkowski core is proposed in \cite{Ling:2021olm}. This novel regular BH overcomes the drawback that the Kretschmann scalar curvature is unbounded by the Planck mass density. Consequently, the metric is applicable to BHs with arbitrarily large masses, making it suitable throughout the process of BH evaporation. Especially, these BHs can reproduce the geometry of Bardeen BH, Hayward BH, or Frolov BH at large scales, depending on the chosen potential form. Furthermore, the authors in \cite{Ling:2022vrv,Zeng:2022yrm, Zeng:2023fqy} explore the features of the photon sphere and the extremal stable circular orbit (ESCO) for massive particles, the accretion disk, and the shadow over this novel regular BH. These analyses contribute to the differentiation of this novel BH from others through astronomical observation. In this paper, we aim to study the properties of the quasinormal modes (QNMs)  over this novel regular BH.
	
QNMs are the oscillatory patterns that emerge when a BH is perturbed, offering insights into the stability and spacetime properties of the BH \cite{Berti:2004md}. The frequencies and damping times associated with these oscillations serve as key characteristics of QNMs, providing essential information about the internal structure and properties of the BH. Departures from GR are anticipated to leave a discernible mark on the QNM spectra, serving as a distinct means of investigating quantum gravity effects \cite{Berti:2005ys,Berti:2018vdi,Fu:2022cul,Fu:2023drp,Gong:2023ghh,Moura:2021eln,Moura:2021nuh,Moura:2022gqm,Lin:2024ubg,Ghosh:2022gka}. 
	
Recently, the analysis of gravitational wave (GW) data from LIGO/Virgo has provided evidence supporting the existence of overtone patterns \cite{Isi:2020gwy, LIGOScientific:2020tif}. Theoretically, overtones offer more precise information about the mass and spin of the remnant BH compared to results derived solely from the fundamental mode \cite{Giesler:2019uxc}. Furthermore, a number of researches  have also highlighted the significance of overtones in effectively modeling the ringdown phase \cite{Giesler:2019uxc,Oshita:2021iyn,Forteza:2021wfq,Oshita:2022pkc}. These findings challenge the dominant notion that the fundamental mode predominantly dictates the signal and suggest that the quasinormal ringing may start earlier than previously expected.
	
Moreover, studies have discovered that the characteristics of high overtones are linked to the geometric properties around the event horizon of a BH \cite{Konoplya:2022pbc}. It is widely recognized that modifications to Einstein's theory will definitely result in significant deformations of the BH's geometry near the event horizon. While the fundamental mode remains relatively unaffected by alterations in the event horizon's geometry, even minor changes in this region can exert a substantial influence on the first few overtones, as observed in \cite{Konoplya:2022pbc}. This highlights the potential to explore the geometry of the BH's event horizon through the investigation of overtones. Such findings motivate further exploration of overtones in various BH scenarios \cite{Konoplya:2022hll,Konoplya:2022iyn,Konoplya:2023aph,Konoplya:2023ahd}.
	
In this paper, our objective is to explore the properties of the QNMs, including both the fundamental mode and the high overtones, for regular BHs with a Minkowski core and sub-Planck curvature. Then, we will compare the QNM properties with those of regular BHs featuring a dS core, as well as loop quantum gravity  corrected (LQG-corrected) BH. The paper is structured as follows. Section \ref{rbh-gbh} provides a concise overview of the regular BH with a Minkowski core and sub-Planck curvature, along with the perturbations of the scalar field over this regular BH. In particular, we delve into the properties of the effective potential, which is intricately connected to the characteristics of the QNMs.
In Section \ref{reular-Quasinormal}, we study the properties of quasinormal
frequencies (QNFs), including both the fundamental mode and the overtones. Additionally, a comparison is made among the QNFs of the regular BH we studied, Hayward BH, and LQG-corrected BH in Section \ref{reular-Hay-LQG}. We summarize our findings in Section \ref{conclusion}. The numerical approaches and error analysis of QNMs are briefly elaborated in Appendix \ref{method}.
	
\section{Scalar field over the regular black hole}\label{rbh-gbh}
The metric of the regular BH with Minkowski core and sub-Planckian Kretschmann scalar curvature has the following form \cite{Ling:2021olm}
\begin{eqnarray} \label{metric}
ds^2=-f(r)dt^2+\frac{1}{f(r)}dr^2+r^2d\Omega^2\,.
\end{eqnarray}
Here, the function $f(r)$ is given by: 
\begin{eqnarray} \label{fr}
f(r)=1+2 P(r)\,,
\end{eqnarray}
where $P(r)$ represents the modified Newton potential defined as:
\begin{eqnarray} \label{Pr}
P(r)=-\frac{M}{r}e^{\frac{-\alpha_{0}M^x}{r^c}}\,.
\end{eqnarray}
In this equation, $M$ represents the mass of the BH, while $\alpha_0$, $x$, and $c$ are dimensionless parameters. Specifically, $\alpha_0$ represents the degree of deviation from the Newton potential. We would like to emphasize that the aforementioned modified Newton potential exhibits an exponentially suppressing form, playing a crucial role in the construction of a regular black hole with an asymptotically Minkowski core. When $\alpha_{0}=0$, the modified Newton potential given by Eq.\eqref{Pr} becomes equivalent to the standard Newton potential. As a consequence, the regular BH is transformed into the Schwarzschild BH (SS-BH). While the parameters $x$ and $c$ characterize the unique features of the regular BH with sub-Planckian curvature. To guarantee that the existence of a BH horizon and the scalar curvature to be sub-Planckian, both the parameters $x$ and $c$ must satisfy the conditions $c\geq x \geq c/3$ and $c \geq 2$. 
Without loss of generality, we set $x=1$ and $c=3$ throughout the paper.
	
\begin{figure}[ht]
	\centering{
		\includegraphics[width=9cm]{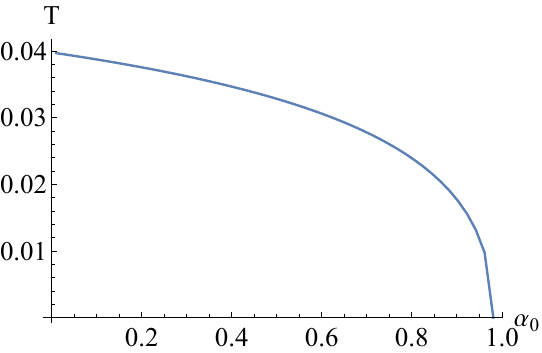}
		\caption{The Hawking temperature as a function of the deviation parameter $\alpha_0$. Here, we have set $r_h=1$ and $M=1$.}
		\label{Hawking_T}
	}
\end{figure}

Subsequently, we can compute the Hawking temperature of the regular BH, which is expressed as follows:
\begin{eqnarray} \label{Tem}
T=\frac{f'(r_h)}{4\pi}=\frac{e^{-\frac{M \alpha_0}{r_h^3}}M(r_h^3-3M \alpha_0)}{2\pi r_h^5}\,, 
\end{eqnarray}
where $r_h$ is the event horizon of the BH. In the extremal case, i.e., when $T=0$, the deviation parameter $\alpha_{0}$ should satisfy the following condition:
\begin{eqnarray} 
\alpha_0\leq\frac{8M^2}{3e}\,.
\end{eqnarray}
Note that the $e$ in the above equation is just the natural constant, also known as the Euler's number.
To illustrate the above relationship, we present the Hawking temperature as a function of the deviation parameter $\alpha_0$ in Fig.\ref{Hawking_T} for visualization.
	
We proceed to study how this regular BH responds to the perturbations of a probing massless scalar field $\Phi$. We can describe the dynamics of the scalar field using the Klein-Gordon (KG) equation:
\begin{eqnarray}
\frac{1}{\sqrt{-g}}\partial_\nu(g^{\mu\nu}\sqrt{-g}\partial_\mu\Phi)=0\,.
\label{scalar_eq}
\end{eqnarray}
Due to the spherical symmetry of spacetime we are investigating, we may utilize the technique of separating variables using spherical harmonics:
\begin{eqnarray}\label{separate}
\Phi(t,r,\theta,\phi)=\sum_{l,m}Y_{l,m}(\theta,\phi)\frac{\Psi_{l,m}(t,r)}{r}\,.
\end{eqnarray}
Here, $Y_{l,m}(\theta,\phi)$ is the spherical harmonics, with $l$ and $m$ representing the angular and azimuthal quantum numbers, respectively. This transformation reformulates the Klein-Gordon equation into a unified Schr$\ddot{o}$dinger-like form:
\begin{eqnarray}\label{Sch_like_eq}
-\frac{\partial^2\Psi}{\partial t^2}+\frac{\partial^2\Psi}{\partial r_{\ast}^2}   - V_{\text{eff}}\Psi = 0\,,
\end{eqnarray}
where $r_*$ is the tortoise coordinate related to $r$ as $dr_{*}/dr=1/f(r)$. For a given $l$ and $m$, we have simplified the notation by removing the subscript from $\Psi_{l,m}(t,r)$ and denoted it as $\Psi$.
In addition, the effective potential is expressed as follows in the equation above:
\begin{eqnarray}\label{V_eff1}
V_{\text{eff}}=f(r)\frac{l(l+1)}{r^2}+\frac{f(r)f'(r)}{r}\,.
\end{eqnarray}
The angular quantum number $l$ takes the values of $0, 1, \ldots$.
	
\begin{figure}[ht]
	\centering{
		\includegraphics[width=7.2cm]{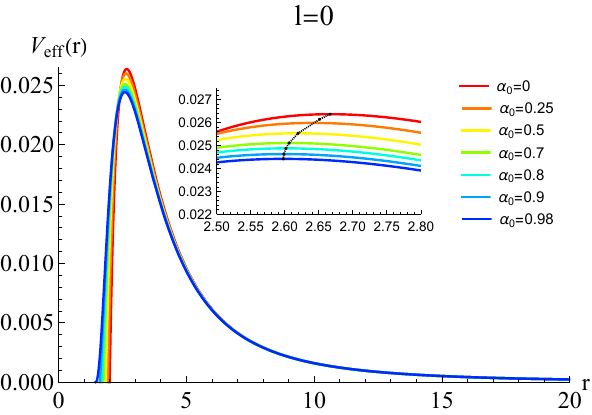}\hspace{0.5cm}
		\includegraphics[width=7.2cm]{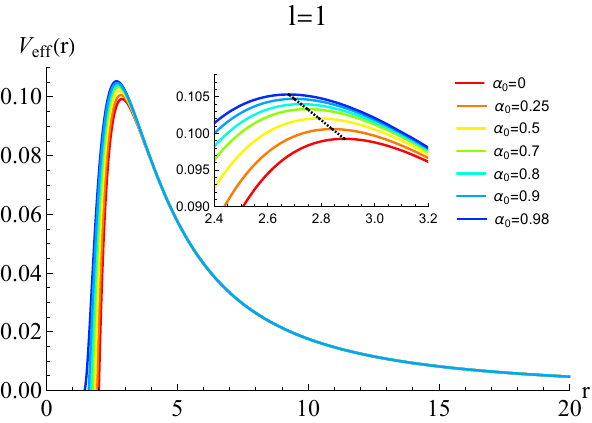}\
		\caption{The effective potential $V_{\text{eff}}(r)$ of the scalar field with different deviation parameters $\alpha_0$. The left panel is for $l=0$, while the right panel is for $l=1$. The black dotted line represents the maximum of the effective potential $V_{\text{eff}}(r)$.}
		\label{V_eff_sca}
	}
\end{figure}

Fig.\ref{V_eff_sca} depicts the behavior of the effective potential $V_{\text{eff}}$ as a function of the coordinate $r$ for the scalar field, showcasing various deviation parameters $\alpha_{0}$. We can find the effective potential is always positive, indicating that the system is stable under the scalar field perturbation. In the case of $l=0$, the first term disappears, leaving only the second term. In this scenario, an increase in the deviation parameter $\alpha_{0}$ leads to a decrease in the peak of the potential, as evident in the inset shown in Fig.\ref{V_eff_sca}.
When $l>0$, the first term of the effective potential becomes non-zero. In this scenario, it is observed that as the deviation parameter $\alpha_{0}$ increases, the peak of the potential rises (the right panel in Fig.\ref{V_eff_sca}), which demonstrates a contrasting trend compared to the $l=0$ case.
	
\section{Quasinormal modes} \label{reular-Quasinormal}
	
In this section, we will thoroughly investigate the properties of the QNMs associated with this regular BH. To determine the QNM spectra, it is imperative to apply the following boundary conditions:
\begin{eqnarray}\label{boundary}
\Psi(r_{\ast})\propto e^{\pm i\omega r_{\ast}}\,,~~~ r_{\ast}\rightarrow\pm\infty\,.
\end{eqnarray}
Here, the minus sign and the plus sign correspond to purely ingoing waves near the event horizon and purely outgoing waves at the boundary, respectively. Consequently, there are no waves emanating from either of the boundaries. This implies that a response in perturbations is only detected when the source of perturbations ceases its action. In other words, proper oscillation frequencies are observed once the perturbation source is no longer active.
	
Several methods have been developed for determining QNMs, including the Wentzel-Kramers-Brillouin(WKB) method \cite{1985ApJ-291L33S,Iyer:1986np,Guinn:1989bn,Konoplya:2004ip,Konoplya:2003ii,Matyjasek:2017psv}, the Horowitz-Hubeny method \cite{Horowitz:1999jd}, the continued fraction method \cite{Leaver:1985ax}, the asymptotic iteration method \cite{Ciftci:2005xn,Cho:2009cj,Cho:2011sf}, the pseudo-spectral (PS) method \cite{Boyd:Chebyshev,Jansen:2017oag}, and others. Among these, the WKB method stands out as an extremely efficient approach. The WKB method has undergone significant development from 1st to 13th order to determine the QNFs \cite{Iyer:1986np,Guinn:1989bn,Konoplya:2004ip,Konoplya:2003ii,Matyjasek:2017psv,Konoplya:2019hlu}. It is important to highlight that, when employing WKB methods, a higher order does not automatically ensure higher accuracy in the results. Through meticulous error analysis, we have observed that the results obtained using the ninth-order WKB approximation in this study exhibit a minimum error (refer to the Appendix \ref{method}). Consequently, for our investigation, we will utilize the ninth-order WKB approximation to determine the QNFs.
	
However, the effectiveness of WKB method loses when seeking higher overtones, particularly in the case of $n>l$. Table \ref{tab_scalar_v3} displays the QNFs for the fundamental modes with various deviation parameter $\alpha_0$ at $l=0$ and $l=1$, using WKB and PS methods. It is noteworthy that the difference between the WKB and PS methods is consistently less than a thousandth. This trend holds true even for $l>n$, as demonstrated in Tables \ref{tab_scalar_v4}, \ref{tab_scalar_v5} and \ref{tab_scalar_v6}. But with an increase in the value of $n$, this difference becomes more evident. This difference can be evaluated by the relative error, defined as:
\begin{eqnarray}
\varepsilon_1&=\left |\frac{\omega_{PS}-\omega_{WKB}}{\omega_{PS}}\right | \times 100\%\,.
	\label{error}
\end{eqnarray}
Here, we assume that the results produced from the PS approach are precise. Clearly, in the case of $l=0$, the relative error has reached two percent even for the fundamental mode (refer to Table \ref{tab_scalar_v4}). However, for $l=1$, the relative error is one percent for $n=3$. Meanwhile, for $l=2$, we can calculate the QNF with $n=5$ with a relative error of one percent.
	
Therefore, in this paper, we primarily utilize the WKB method as a cross-check, specifically for determining fundamental modes. For higher overtones, we commonly resort to the PS method—a fully numerical approach renowned for its power in determining QNFs \cite{Jansen:2017oag,Wu:2018vlj,Fu:2018yqx,Xiong:2021cth,Liu:2021fzr,Liu:2021zmi,Jaramillo:2020tuu,Jaramillo:2021tmt,Destounis:2021lum,Fu:2022cul,Fu:2023drp}. Nevertheless, it is essential to note that, like many numerical methods, the PS method may be time-consuming compared to the semi-analytic WKB method. Therefore, when employing the PS method, a thoughtful balance between computational time and desired accuracy is crucial. For a concise introduction to the WKB method and the PS method, please refer to the Appendix \ref{method}.
	
\begin{table}[H]
	\centering
		\begin{tabular}{c|cc|cc}
		\hline \hline &  \multicolumn{2}{c}{\text { $\omega_{WKB}$ }} & \multicolumn{2}{c}{\text {$\omega_{PS}$  }} \\ \hline
		$\alpha_0$ & $l=0$ & $l=1$ & $l=0$ & $l=1$ \\
		\hline 0 & 0.110313-0.104962i & 0.292935-0.097655i & 0.110455-0.104896i & 0.292936-0.097660i \\
		10/52  & 0.112430-0.103153i & 0.295307-0.095988i & 0.111943-0.102876i & 0.295310-0.095994i \\
		20/52  & 0.113303-0.100102i & 0.297754-0.094006i & 0.113316-0.100303i & 0.297757-0.093989i \\
		30/52  & 0.114104-0.096353i & 0.300215-0.091542i & 0.113927-0.092471i & 0.300207-0.091534i \\
		40/52  & 0.113168-0.091982i & 0.302474-0.088522i & 0.112306-0.090224i & 0.302476-0.088519i \\
		45/52  & 0.111758-0.090069i & 0.303425-0.086816i & 0.111252-0.089791i & 0.303430-0.086807i \\
		50/52  & 0.110099-0.089600i & 0.304205-0.085016i & 0.110302-0.089928i & 0.304217-0.085026i \\
		8/3e   & 0.109905-0.089559i & 0.304357-0.084659i & 0.110114-0.089874i & 0.304358-0.084662i \\
		\hline \hline
		\end{tabular}\\
		\caption{The fundamental modes of the regular BH are presented for various $\alpha_0$ with $l=0$ and $l=1$. Left column is for the WKB method, whereas the right column is for the PS method.}\label{tab_scalar_v3}
\end{table} 
	
\begin{table}[H]
	\centering
		\begin{tabular}{c|c|cc|c}
			\hline \hline 
			$l$ &$n$ & $\omega_{WKB}$  & $\omega_{PS}$ &   $\varepsilon_1$ \\
			\hline 
			\multirow{3}{0.3cm}{$0$}
			&0 & 0.114129-0.094736i &0.113983-0.098395i &  2.43$\%$ \\
			&1 & 0.074648-0.318230i &0.083923-0.321838i &  2.99$\%$ \\
			&2 & 0.047307-0.594856i &0.039833-1.335483i &  55.4$\%$\\
			\hline \hline
		\end{tabular}\\
	\caption{A comparison of the QNFs between the results obtained through the WKB method and the PS method. Here we have fixed $\alpha_{0}=\frac{1}{2}$ for $l=0$ with various $n$.}\label{tab_scalar_v4}
\end{table}
	
\begin{table}[H]
	\centering
		\begin{tabular}{c|c|cc|c}
			\hline \hline 
			$l$ &$n$ & $\omega_{WKB}$  & $\omega_{PS}$  &   $\varepsilon_1$ \\
			\hline 
			\multirow{5}{0.3cm}{$1$}  
			&0 & 0.299242-0.092591i &0.299235-0.092579i &  0.00444$\%$ \\
			&1 & 0.272806-0.287503i &0.272726-0.287452i &  0.0239$\%$\\
			&2 & 0.235278-0.502697i &0.234530-0.501686i &  0.227$\%$ \\
			&3 & 0.203221-0.730057i &0.194619-0.729233i &  1.14$\%$ \\
			&4 & 0.174490-0.959562i &0.147923-0.965609i &  2.79$\%$ \\
			\hline \hline
		\end{tabular}\\
	\caption{A comparison of the QNFs between the results obtained through the WKB method and the PS method. Here we have fixed $\alpha_{0}=\frac{1}{2}$ for $l=1$ with various $n$.}\label{tab_scalar_v5}
\end{table}
	
\begin{table}[H]
	\centering
		\begin{tabular}{c|c|cc|c}
			\hline \hline 
			$l$&$n$ &  $\omega_{WKB}$  & $\omega_{PS}$  &  $\varepsilon_1$ \\
			\hline 
			\multirow{6}{0.3cm}{$2$}
			&0 & 0.493663-0.092006i &0.493663-0.092004i &  3.98*$10^{-4}$ $\%$\\
			&1 & 0.475664-0.279938i &0.475673-0.279932i &  0.00196 $\%$\\
			&2 & 0.443879-0.478407i &0.443938-0.478263i &  0.0238 $\%$\\
			&3 & 0.405482-0.690520i &0.405132-0.689352i &  0.152 $\%$\\
			&4 & 0.367827-0.914002i &0.364227-0.911147i &  0.468 $\%$\\
			&5 & 0.334292-1.145390i &0.322035-1.140722i &  1.11 $\%$\\
			\hline \hline
		\end{tabular}\\
	\caption{A comparison of the QNFs between the results obtained through the WKB method and the PS method. Here we have fixed $\alpha_{0}=\frac{1}{2}$ for $l=2$ with various $n$.}\label{tab_scalar_v6}
\end{table}
	
	
\begin{figure}[ht]
	\centering{
		\includegraphics[width=7.8cm]{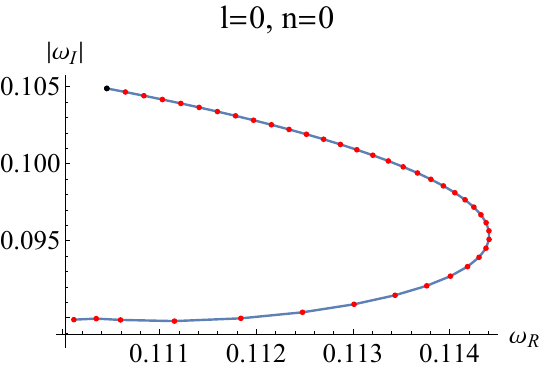}\hspace{6mm}
		\includegraphics[width=7.8cm]{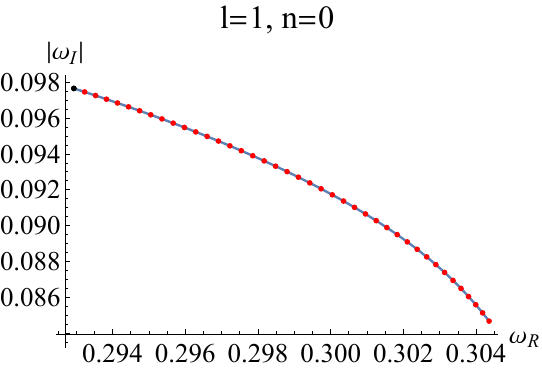}
		\caption{The fundamental modes of the scalar field perturbation over the regular BH are depicted in the left plot for $l=0$ and the right plot for $l=1$. The black dot corresponds to the SS-BH ($\alpha_0=0$), while the red dots correspond to various deviation parameters $\alpha_0$.}
			\label{QNMs_scalar_n0_v0}
	}
\end{figure}
	
Now, we will delve into an investigation of the properties of the QNMs associated with the regular BH. The phase diagram $\left| \omega_I\right|-\omega_R$ illustrating the fundamental modes for this regular BH is presented in Fig.\ref{QNMs_scalar_n0_v0}. The black dot corresponds to the SS-BH with $\alpha_0=0$, while the red dots represent various deviation parameters $\alpha_0$. As the deviation parameter $\alpha_0$ grows, we see that the real part $\omega_R$ shows a distinct non-monotonic pattern (left plot in Fig.\ref{QNMs_scalar_n0_v0}). More precisely, when the parameter $\alpha_0$ grows, indicating a larger departure from the SS-BH, $\omega_R$ first rises, indicating a more pronounced oscillation of the system, and subsequently declines, suggesting a reduction in the amplitude of the oscillation. The turning point of this non-monotonic behavior occurs at approximately $\alpha_0 \approx 13/20$. It is important to mention that when the deviation parameter $\alpha_0$ is increased to a value near to the extremal case, the magnitude of $\omega_R$ becomes less than that of the SS-BH. Simultaneously, when $\alpha_0$ increases, the magnitude of the imaginary part of the QNFs, $\left| \omega_I\right|-\omega_R$, consistently drops, suggesting a reduced damping rate. This finding suggests that the presence of quantum gravity effects leads to decelerated decay modes.
	
However, it is clearly observed in the right plot in Fig.\ref{QNMs_scalar_n0_v0} that the non-monotonic behavior disappears for the case of $l=1$. More precisely, when the parameter $\alpha_0$ grows, both $\omega_R$ and $\omega_I$ increase in a monotonic manner. This indicates that the system exhibits stronger oscillations and slower decay. In addition, we also analyze the fundamental mode for values of $l>1$ and reach a comparable conclusion to that of the situation where $l=1$. A similar observation has been noted in the fundamental modes associated with the LQG-corrected BH \cite{Gong:2023ghh}. This suggests that for the fundamental modes, the influence of the angular number surpasses that of quantum gravity corrections. In the future, it deserves further investigation into the university of the non-monotonic behavior for $l=0$ and exploration of the underlying reasons behind this phenomenon.
\begin{figure}[ht]
		\centering{
			\includegraphics[width=7.8cm]{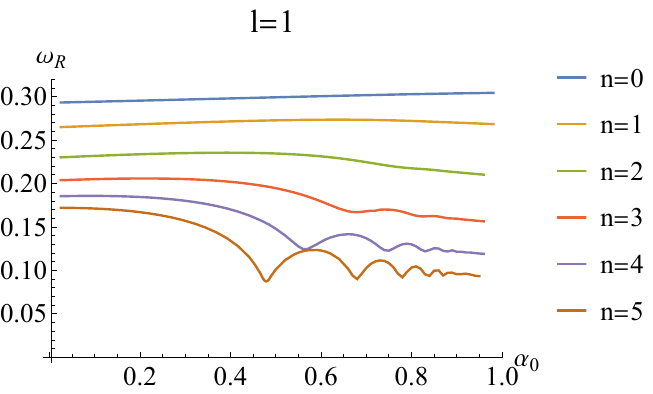}\hspace{6mm}
			\includegraphics[width=7.8cm]{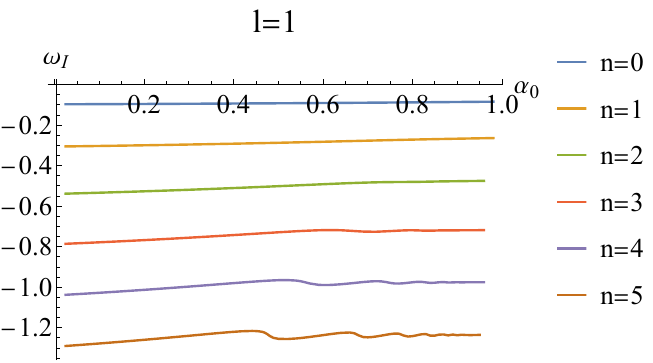}
			\caption{The real part $\omega_R$ is depicted on the left plot, while the right plot illustrates the imaginary part $\omega_I$. Both plots represent the fundamental mode and the first five overtones from top to bottom, with $\alpha_0$ as the varying parameter.}
			\label{QNMs_scalar_fin_v1}
		}
\end{figure}
	
As revealed above, the fundamental modes exhibit interesting properties in the context of this regular BH pattern. However, it is essential to note that upon closer analysis, we observe that the absolute values of the fundamental modes change only by a few percent compared to those of the SS-BH as the deviation parameter $\alpha_0$ increases (refer to Table \ref{tab_scalar_v3}). In Fig.\ref{QNMs_scalar_fin_v1}, the blue curves corresponding to $n=0$ clearly highlight this discrepancy. However, while examining the overtone modes, a notable discrepancy arises in QNFs between the regular BH and the SS-BH.
	
The most crucial feature of the overtone modes is the phenomenon known as the ``outburst of the overtone'', indicating a relatively significant change in the spectra of the overtone modes compared to that of the SS-BH   \cite{Berti:2003zu,Konoplya:2022hll,Konoplya:2022pbc,Konoplya:2022iyn,Konoplya:2023ppx,Konoplya:2023ahd,Fu:2023drp,Gong:2023ghh}. In our current model, such an outburst can also be distinctly observed for $l=1$ beginning from the third overtone (refer to Fig. \ref{QNMs_scalar_fin_v1}). Furthermore, from Fig.\ref{QNMs_scalar_fin_v1}, we also note that as the deviation parameter $\alpha_0$ increases, this regular BH approaches the extremal case and an oscillatory behavior can be clearly observed. This oscillation behavior becomes more pronounced for the higher overtones. Especially, we would like to point out that such oscillation behavior has also been observed in LQG-corrected BH \cite{Fu:2023drp,Gong:2023ghh}. This oscillatory pattern may be associated with the extremal effect. But the reason behind this phenomenon deserves further exploring in the future.
\begin{figure}[ht]
	\centering{
		\includegraphics[width=7.8cm]{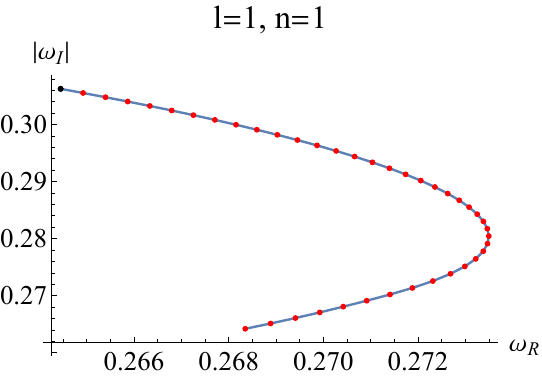}\hspace{6mm}
		\includegraphics[width=7.8cm]{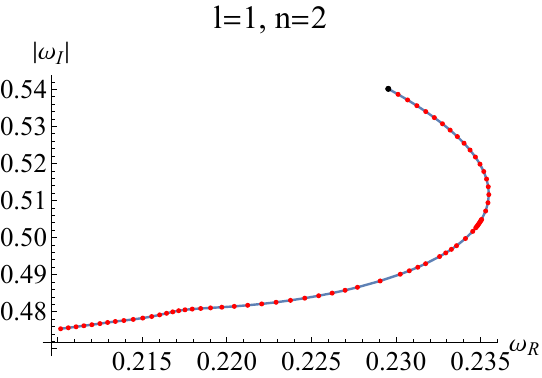}
		\caption{Phase diagram $\left| \omega_I\right|-\omega_R$ for $l=1$ with $n=1$ and $n=2$. The black dot corresponds to the SS-BH ($\alpha_0=0$), while the red dots correspond to various deviation parameters $\alpha_0$.
		}
		\label{QNMs_scalar_ln_v0v1}
	}
\end{figure}
	
Recall that the non-monotonic behavior of QNFs appears in the case of $l=0$ for the fundamental modes, but this non-monotonic pattern disappears for $l\geq 1$. Surprisingly, for the overtones, we observe the reemergence of this non-monotonic behavior (refer to Fig. \ref{QNMs_scalar_ln_v0v1}). Specially, for $l=1$ and $n=2$, as $\alpha_0$ increases, the magnitude of $\omega_R$ becomes smaller than that of the SS-BH, similar to the behavior observed in the fundamental modes with $l=0$. However, we observe that the critical point of $\alpha_0$ at which the magnitude of $\omega_R$ becomes smaller than that of the SS-BH is smaller than that of the fundamental modes with $l=0$.
\begin{figure}[ht]
	\centering{
		\includegraphics[width=5.2cm]{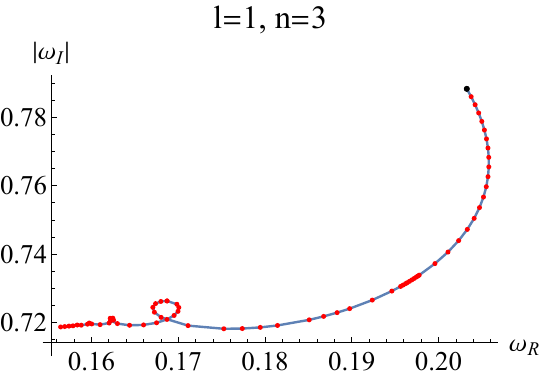}\hspace{2mm}
		\includegraphics[width=5.2cm]{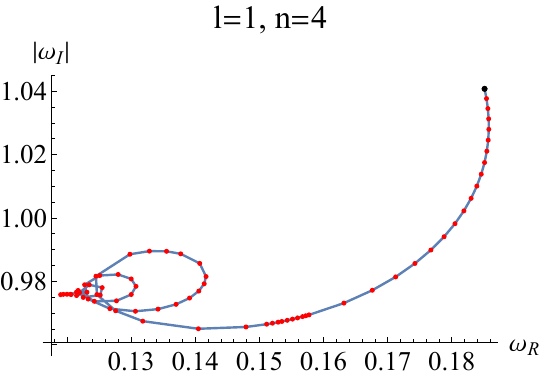}\hspace{2mm}
		\includegraphics[width=5.2cm]{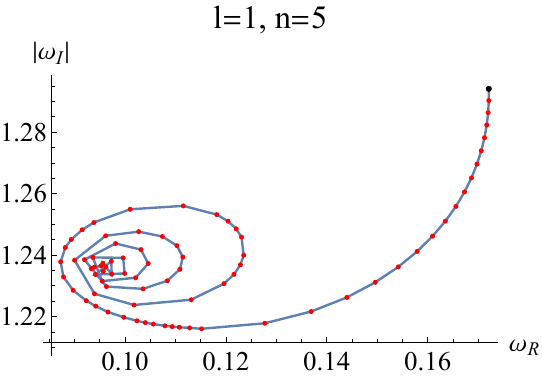}
		\caption{Phase diagram $\left| \omega_I\right|-\omega_R$ for $l=1$ with various overtone modes. The black dot corresponds to the SS-BH ($\alpha_0=0$), while the red dots correspond to various deviation parameters $\alpha_0$.}
		\label{QNMs_scalar_ln_v0}
	}
\end{figure}
\begin{figure}[ht]
	\centering{
		\includegraphics[width=7.8cm]{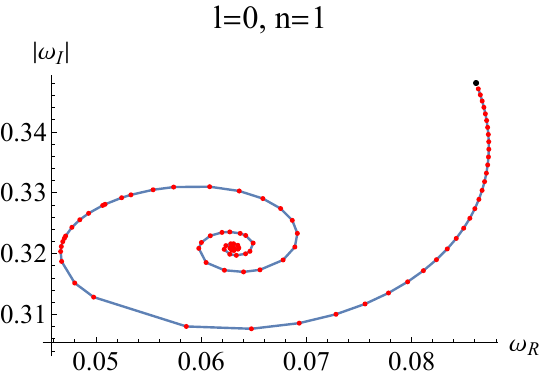}
		\caption{Phase diagram $\left| \omega_I\right|-\omega_R$ for $l=0$ with $n=1$. The black dot corresponds to the SS-BH ($\alpha_0=0$), while the red dots correspond to various deviation parameters $\alpha_0$.}
		\label{QNMs_scalar_fin_v0}
	}
\end{figure}
	
Upon further examination of the phase diagram $\left| \omega_I\right|-\omega_R$ for $l=1$ and higher overtone modes, a spiral-like structure is identified in Fig.\ref{QNMs_scalar_ln_v0}. As the overtone number increases, this spiral-like structure becomes more pronounced. Notably, for $l=0$, a clearly visible spiral-like structure is observed for the first overtone mode (see Fig.\ref{QNMs_scalar_fin_v0}). This spiral-like pattern is, in fact, a manifestation of the oscillatory behaviors observed in Fig.\ref{QNMs_scalar_fin_v1}. Such a spiral-like structure has been observed in the Hayward BH \cite{Konoplya:2022hll} and the LQG-corrected BHs \cite{Fu:2023drp,Gong:2023ghh,Moreira:2023cxy}.
	
\begin{figure}[ht]
	\centering{
		\includegraphics[width=7cm]{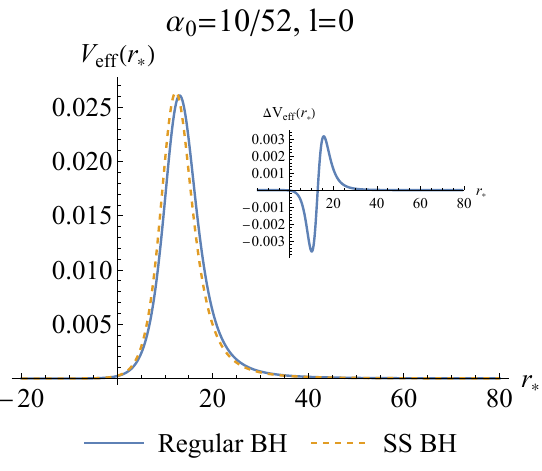}\hspace{6mm}
		\includegraphics[width=7cm]{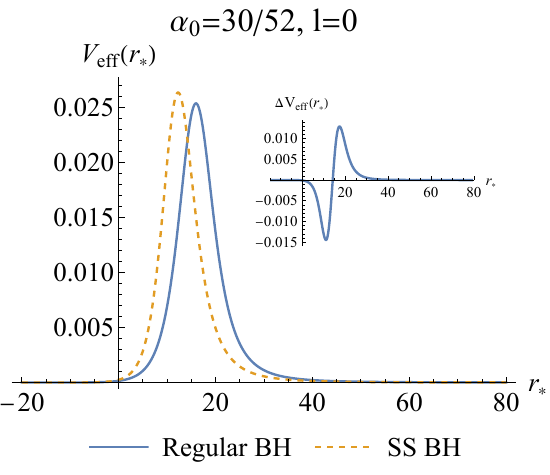}\ \\
		\includegraphics[width=7cm]{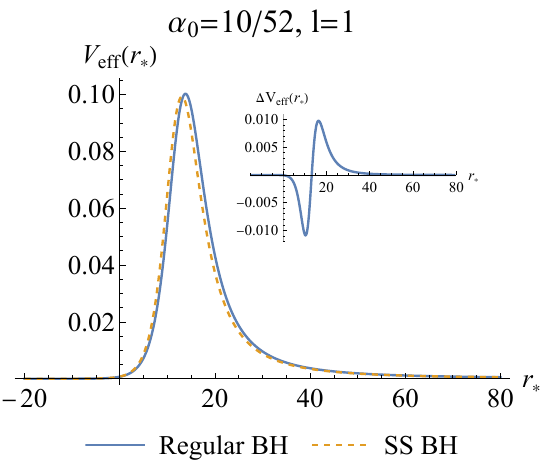}\hspace{6mm}
		\includegraphics[width=7cm]{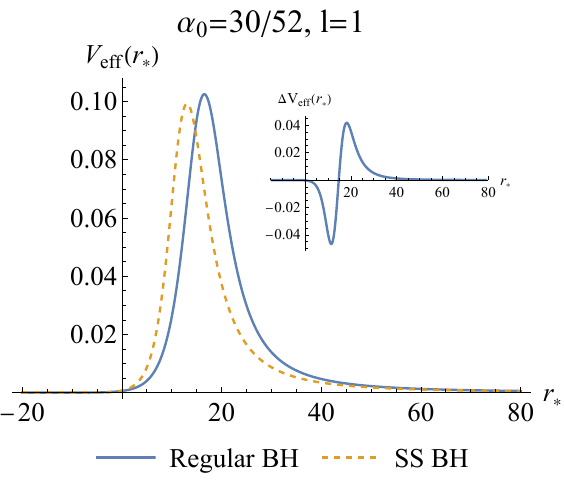}
		\caption{The difference in the effective potentials between the regular BH and the SS-BH for various deviation parameters $\alpha_0$ and angular quantum numbers $l$. 
		}
		\label{QNMs_RVeffv0}
	}
\end{figure}
	
Previous studies have suggested that the outburst in overtones may be associated with subtle differences between the SS-BH and the modified effective SS-BH \cite{Berti:2003zu,Konoplya:2022hll,Konoplya:2022pbc,Konoplya:2022iyn,Konoplya:2023ppx,Konoplya:2023ahd,Fu:2023drp,Gong:2023ghh}. This assertion is further supported in our current model. Fig.\ref{QNMs_RVeffv0} illustrates a comparison of the effective potential between this regular BH and the SS-BH for different values of $\alpha_0$, considering $l=0$ and $l=1$. This comparison is measured by the quantity $\Delta V$, which quantifies the difference in the effective potentials (see the inset in Fig.\ref{QNMs_RVeffv0}). Evidently, there is a minor change in the region surrounding the event horizon. Increasing the deviation parameter $\alpha_0$ causes a more significant change in the effective potential. This change in the effective potential may act as a trigger for the outburst of overtones. Therefore, the deviation parameter $\alpha_0$ significantly influences the outburst of overtones.

\section{A comparison among the QNFs of the regular BH, Hayward BH, and LQG-corrected BH}\label{reular-Hay-LQG}

Throughout this paper, we have employed the values $x=1$ and $c=3$. In this setting, the redshift factor \eqref{metric} of the regular BH can be expanded at large scales, i.e., as $r\to\infty$,
\begin{eqnarray}\label{RLfr}
	f(r)\cong 1-\frac{2M}{r}\left(1-\frac{\alpha_{0}M}{r^3}+\cdots\right)\,.
\end{eqnarray}
If we also expand the redshift function of the Hayward BH at large scales, we observe:
\begin{eqnarray}\label{HLfr}
	f(r)=1-\frac{2Mr^2}{r^3+M \alpha_{0}}\cong 1-\frac{2M}{r}\left(1-\frac{\alpha_{0}M}{r^3}+\cdots\right)\,.
\end{eqnarray}
Remarkably, we find that the large-scale behavior of the Hayward BH is identical to that of the regular BH with $x=1$ and $c=3$. This similarity has been pointed out in \cite{Ling:2021olm}. Interestingly, a LQG-corrected BH within the entire spacetime has the following form \cite{Lewandowski:2022zce}:
\begin{eqnarray}\label{RLfr}
	f(r)=1-\frac{2M}{r}+\frac{2\alpha_{0}M^2}{r^4}\,,
\end{eqnarray}
which is just the large-scale behavior of the Hayward BH and the regular BH with $x=1$ and $c=3$.
	
\begin{figure}[ht]
		\centering{
			\includegraphics[width=7.76cm]{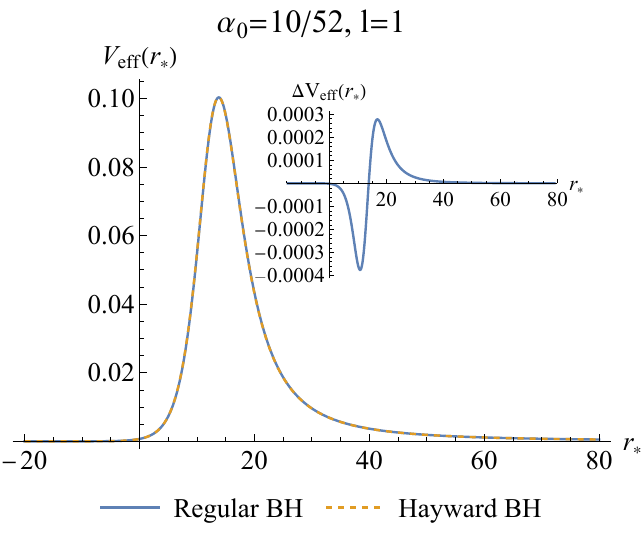}\hspace{6mm}
			\includegraphics[width=7.7cm]{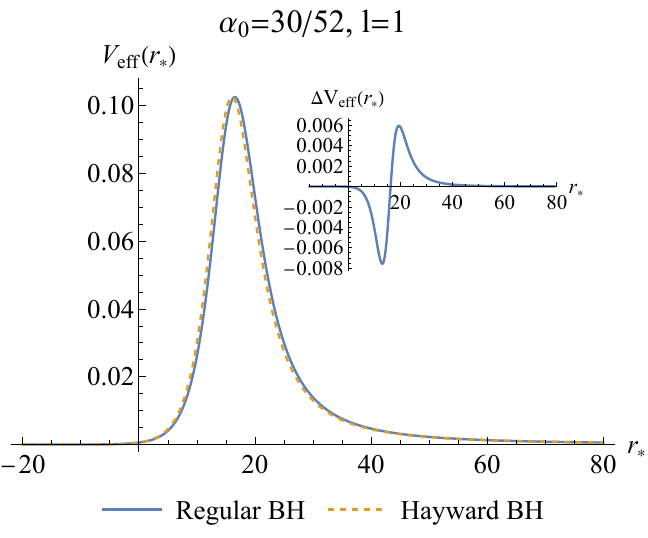}
			\caption{The difference in the effective potentials between the regular BH and the Hayward BH for various deviation parameters $\alpha_0$. Here we have fixed the angular quantum number $l=1$.
			}
			\label{QNMs_RVeff-H}
		}
\end{figure}
\begin{figure}[ht]
		\centering{
			\includegraphics[width=7.7cm]{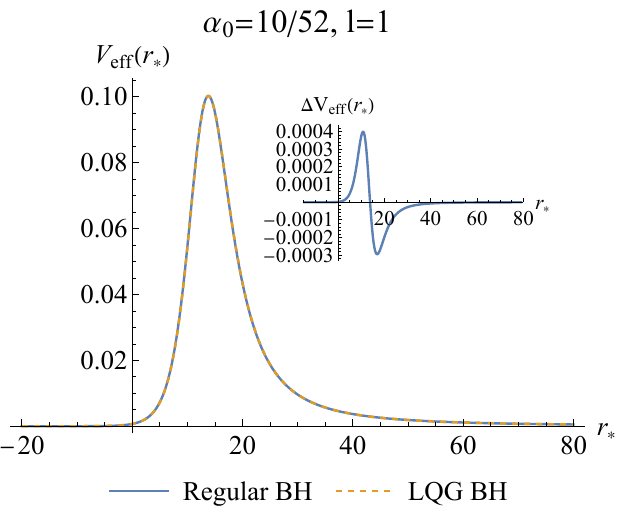}\hspace{6mm}
			\includegraphics[width=7.6cm]{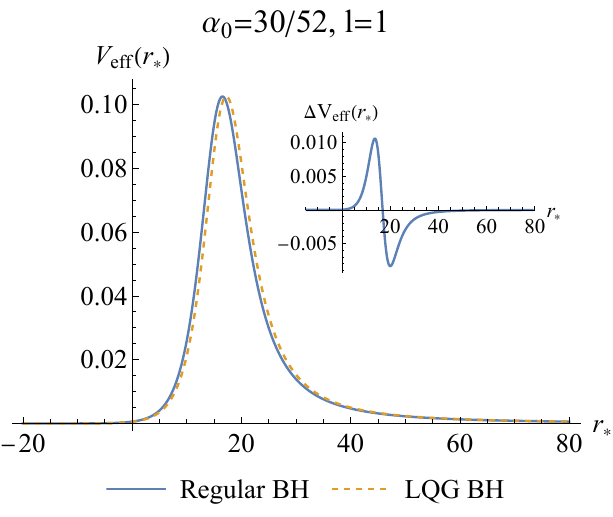}
			\caption{The difference in the effective potentials between the regular BH and the LQG-corrected BH for various deviation parameters $\alpha_0$. Here we have fixed the angular quantum number $l=1$.}
			\label{QNMs_LVeff-L}
		}
\end{figure}
Next, we will compare the properties of the QNFs of the regular BH studied in this paper with those of a Hayward BH and a LQG-corrected BH. This comparative analysis seeks to evaluate the viability of QNFs as a reliable probe for discerning among these distinct models. To begin with, we illustrate the difference in the effective potentials between the Hayward BH and the regular BH in Fig.\ref{QNMs_RVeff-H}, and likewise, the difference between the effective potentials of the LQG-corrected BH and the regular BH in Fig.\ref{QNMs_LVeff-L}. For a small deviation parameter $\alpha_0$, it appears that all three models have identical effective potentials, as seen in the left plots of Figs.\ref{QNMs_RVeff-H} and \ref{QNMs_LVeff-L}. However, when doing meticulous investigation, we observe nuanced distinctions among them. Particularly, the difference is obvious in the vicinity of the event horizon (refer to the insets in the left plots in Figs.\ref{QNMs_RVeff-H} and \ref{QNMs_LVeff-L}). The disparity becomes more evident with a larger deviation parameter $\alpha_0$, as see the right plots in Figs.\ref{QNMs_RVeff-H} and \ref{QNMs_LVeff-L} for $\alpha_0=30/52$. 
	
It is expected that subtle differences in the vicinity of the event horizon can be distinguished by the QNFs, as discussed earlier. To investigate this, we calculate the QNFs for these three black hole models with the same parameter $\alpha_0$. Furthermore, we quantify the difference between the regular BH and either the Hayward BH or the LQG-corrected BH using the following definition: 
\begin{eqnarray}
	\varepsilon_2=\left |\frac{\omega_{Regular}-\omega_{\#}}{\omega_{Regular}}\right | \times 100\%\,,
	\label{errorv2}
\end{eqnarray}
where $\omega_{Regular}$ represents the QNF of the regular BH, and $\omega_{\#}$ denotes the QNF of either the Hayward BH or the LQG-corrected BH.
The results are presented in Table \ref{tab_regular_v0} for $\alpha_0=10/52$ and Table \ref{tab_regular_v1} for $\alpha_0=30/52$.

\begin{table}[H]
		\centering
		\begin{tabular}{c|c|c|cc|cc}
			\hline \hline 
			$\alpha_0$&$n$ & $\omega_{Regular}$ & $\omega_{Hayward}$ &  $\varepsilon_2$ & $\omega_{LQG}$  &  $\varepsilon_2$ \\
			\hline 
			\multirow{6}{0.5cm}{$\frac{10}{52}$}
			&0 & 0.29531-0.09599i & 0.29531-0.09603i &0.0129$\%$ & 0.29531-0.09596i & 0.00966 $\%$\\
			&1  & 0.26808-0.30010i & 0.26811-0.30024i & 0.0356$\%$ & 0.26805-0.29996i & 0.0356 $\%$\\
			&2  & 0.23356-0.52754i & 0.23373-0.52784i & 0.0598$\%$ & 0.23337-0.52723i & 0.0630 $\%$\\
			&3  & 0.20578-0.76874i & 0.20623-0.76922i & 0.0827$\%$ & 0.20527-0.76824i & 0.0897 $\%$\\
			&4  & 0.18468-1.01443i & 0.18558-1.01509i & 0.108$\%$ & 0.18368-1.01374i & 0.118$\%$ \\
			&5  & 0.16750-1.26128i & 0.16903-1.26212i & 0.137$\%$ & 0.16575-1.26043i & 0.153 $\%$\\
			\hline \hline
		\end{tabular}\\
		\caption{A comparison of the QNFs between the regular BH and either the Hayward BH or the LQG-corrected BH. Here we have fixed $\alpha_{0}=\frac{10}{52}$ for $l=1$ with various $n$.}\label{tab_regular_v0}
\end{table}
\begin{table}[H]
	\centering
\begin{tabular}{c|c|c|cc|cc}
			\hline \hline 
			$\alpha_0$&$n$ & $\omega_{Regular}$ & $\omega_{Hayward}$ &  $\varepsilon_2$ & $\omega_{LQG}$  &  $\varepsilon_2$ \\
			\hline 
			\multirow{6}{0.5cm}{$\frac{30}{52}$}
			&0 & 0.30021-0.09154i & 0.30027-0.09195i & 0.131$\%$ & 0.30007-0.09108i& 0.153$\%$\\
			&1 & 0.27333-0.28363i & 0.27439-0.28522i & 0.485$\%$ & 0.27171-0.28202i & 0.580$\%$\\
			&2 & 0.23230-0.49423i & 0.23699-0.49719i & 1.02$\%$ & 0.22516-0.49256i& 1.34 $\%$\\
			&3 & 0.18388-0.72019i & 0.19819-0.72189i &1.94$\%$ &0.16671-0.73513i &3.06 $\%$\\
			&4 & 0.12592-0.98492i & 0.15475-0.95352i &4.29$\%$&  0.14700-0.99454i &2.33$\%$\\
			&5 & 0.12360-1.24380i & 0.08530-1.21242i &3.96$\%$& 0.11149-1.23661i &1.13$\%$\\
			\hline \hline
		\end{tabular}\\
		\caption{A comparison of the QNFs between the regular BH and either the Hayward BH or the LQG-corrected BH. Here we have fixed $\alpha_{0}=\frac{30}{52}$ for $l=1$ with various $n$. }\label{tab_regular_v1}
\end{table}
	
From Table \ref{tab_regular_v0}, we observe that for $\alpha_0=10/52$, the discrepancy in the fundamental modes among the three BH models is merely $0.01$ percent. However, as the overtone number increases, the disparity grows, reaching approximately $0.15$ percent for $n=5$. Increasing the deviation parameter to $\alpha_0=30/52$, we note that the difference among these models reaches approximately $0.15$ percent even for the fundamental modes, particularly escalating to around $4.3$ percent for some higher overtones. These trends can be characterized as the outburst of overtones in the regular BH when compared to the Hayward BH and the LQG-corrected BH. In conclusion, the QNFs undeniably serve as a reliable tool for distinguishing these distinct models.

\section{ CONCLUSION AND DISCUSSION}\label{conclusion}
	
In this paper, we investigate the properties of the QNMs of a regular BH characterized by a Minkowski core and sub-Planckian curvature. Our specific focus is on the instance of this regular BH with fixed parameters $x=1$ and $c=3$, where the only free parameter is the deviation parameter $\alpha_0$. This regular BH shares identical large-scale behavior with the Hayward BH, and notably, the large-scale behavior of both these BHs corresponds to the geometry of the LQG-corrected BH in \cite{Lewandowski:2022zce}.
	
Owing to the influence of quantum gravity, the fundamental modes display a non-monotonic trend concerning the deviation parameter $\alpha_0$ when $l=0$. However, this non-monotonic pattern diminishes for $l\geq 1$, indicating that the impact of the angular quantum number supersedes that of the quantum gravity effect. This non-monotonic behavior reemerges in the modes of higher overtones.
	
Furthermore, we investigate the properties of the overtones. The most crucial feature of the overtone modes is the outburst of the overtones compared to that of SS-BH. This effect has been noticed in several geometric situations other than the SS-BH, such as the Reissner-Nordström BH, Bardeen BH, higher-derivative gravity model, and certain quantum-corrected BHs. The outburst of the overtones is expected to detect the effects deviating from the SS-BH, particularly the impact of quantum gravity. In addition, we also analyze and contrast the characteristics of the QNFs of the regular BH with those of the Hayward BH and the LQG-corrected BH. We observe that the disparity in the QNFs between the regular BH and the Hayward BH or the LQG-corrected BH is roughly $4.3$ percent for certain high overtones. Such a relatively large change of the overtone modes in the regular BH, as compared to the Hayward BH and the LQG-corrected BH, may also be referred to as the outburst of the overtones. Hence, the QNFs unquestionably function as a dependable instrument for discerning various BH types.
	
The discovery of GWs has provided a novel means of detecting the effects of quantum gravity and distinguishing between different theories of BHs. The next crucial step is the investigation of gravitational perturbations and the analysis of their QNMs. By employing QNMs, encompassing both the fundamental mode and high overtone modes, one can achieve an accurate alignment with gravitational waves emitted during the ringdown phase. As a result, by analyzing QNMs and GW signals during the ringdown phase, we may evaluate quantum gravity effects and differentiate between various BH models. The investigation of these theories and experimental findings has the potential to enhance our comprehension and study of quantum gravity. It advances our understanding of BH physics and GW astronomy while also uncovering fundamental natural laws.

\acknowledgments
	
This work is supported by National Key R$\&$D Program of China (No. 2020YFC2201400), the Natural Science Foundation of China under Grants Nos. 12375055, 12347159, 12275079 and 12035005, the Postgraduate Scientific Research Innovation Project of Hunan Province under Grant No. CX20220509, the Postgraduate Research $\&$ Practice Innovation Program of Jiangsu Province under Grant No. KYCX22$\_$3451.
	
\appendix
	
\section{The numerical methods}\label{method}
	
This appendix provides a brief introduction of the numerical approaches employed in our work, namely the WKB method and the PS method. 
	
\subsection{WKB Method}\label{appendix-A}
	
The WKB method serves as an efficient tool for solving the Schr\"{o}dinger-like equation \eqref{Sch_like_eq}. Over the years, significant advancements in the WKB method, ranging from 1st to 13th order, have been made to determine the QNFs \cite{Iyer:1986np,Guinn:1989bn,Konoplya:2004ip,Konoplya:2003ii,Matyjasek:2017psv,Konoplya:2019hlu}. It is crucial to note that a higher order in the WKB method does not inherently guarantee higher accuracy in the results. In this subsection, we present a succinct overview of the WKB method. We will also perform an error analysis for various orders of the WKB method when determining the QNFs for our current model, aiming to identify the most suitable order for handling our specific model.
	
Typically, the higher-order WKB formula can be formulated as follows:
\begin{eqnarray}\label{WKB_formula}
		\omega^2=V_0 + A_2(\mathcal{K}^2) + A_4(\mathcal{K}^2) +....- i\mathcal{K}\sqrt{-2V_2}\left[1+A_3(\mathcal{K}^2)+A_5(\mathcal{K}^2)+....\right]\,.
\end{eqnarray}
In this context, $V_0$ denotes the maximum of the effective potential, $V_2$ stands for the second-order derivative of the effective potential with respect to $r$ at its maximum position, and $\mathcal{K}$ assumes a value that is a half-integer. $A_k(\mathcal{K}^2)$ is the correction term of order $k$, comprising polynomials in $\mathcal{K}^2$ with coefficients as rational numbers. This correction term depends on the higher derivative of the effective potential $V_{\text{eff}}$ at its maximum position. For more details, please refer to \cite{Konoplya:2019hlu,Konoplya:2011qq,Churilova:2019cyt}.

\begin{figure}[ht]
	\centering{
		\includegraphics[width=7.8cm]{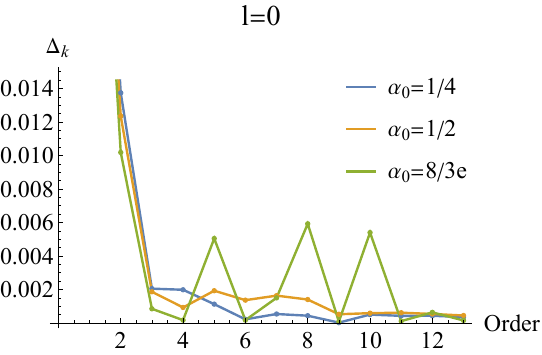}\hspace{6mm}
		\includegraphics[width=7.8cm]{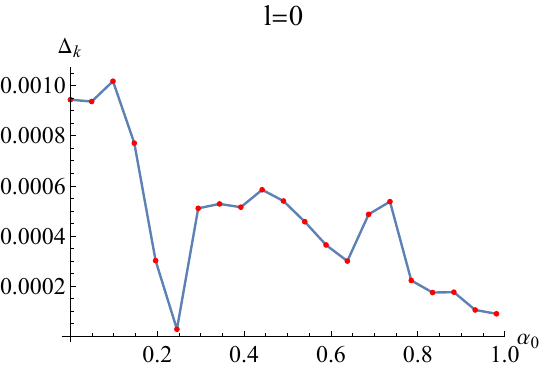}
		\caption{The error evaluation for the fundamental modes.  The error estimation function $\Delta_{k}$, illustrating its dependence on the WKB order for various deviation parameters $\alpha_0$, is presented in the left plot. Simultaneously, the right plot illustrates its dependence on $\alpha_0$ for the ninth-order WKB approximation.}
		\label{WKB_error}}
\end{figure}

To assess the most suitable order of the WKB method for our model, we adopt the error estimation function $\Delta_k$ in the work \cite{Konoplya:2019hlu}:
\begin{eqnarray}\label{error_estimate}
		\Delta_k=\frac{|\omega_{k+1}-\omega_{k-1}|}{2}\,.
\end{eqnarray}
The left plot in Fig.\ref{WKB_error} illustrates the error estimation function $\Delta_{k}$ as a function of the WKB order for the fundamental mode, considering different quantum-corrected parameters $\alpha_0$. Our analysis
shows that the ninth-order WKB approximation produces a comparatively tiny $\Delta_{k}$ in the model. In addition, we illustrate the relationship between $\Delta_{k}$ and $\alpha_0$ for the ninth-order WKB approximation, clearly showing that it is consistently less than $0.105\%$. These findings lead us to the conclusion that the ninth-order WKB approximation provides a superior degree of accuracy when compared to WKB approximations of other orders in our model. Therefore, we will follow the ninth-order WKB technique to calculate the QNFs in this case.
	
\subsection{Pseudo-spectral method}\label{appendix-B}
	
The PS method, which is a fully numerical technique, is highly effective in determining QNFs. It has been widely utilized in various studies \cite{Jansen:2017oag,Wu:2018vlj,Fu:2018yqx,Xiong:2021cth,Liu:2021fzr,Liu:2021zmi,Jaramillo:2020tuu,Jaramillo:2021tmt,Destounis:2021lum,Fu:2022cul,Fu:2023drp}, particularly for identifying high overtone modes \cite{Fu:2023drp}. The fundamental principle of the PS approach is to discretize the differential equations and subsequently solve the resultant generalized eigenvalue equations. Specifically, we substitute the continuous variables with a discrete collection of collocation points known as the grid points, and we expand the functions using specific basis functions referred to as cardinal functions. Typically, the Chebyshev grids and Lagrange cardinal functions are employed:
\begin{eqnarray}
	\label{Cg-Lcf}
	x_i=\cos\left( \frac{i}{N}\pi\right) \,, \ \ C_j(x)=\prod_{i=0,i\neq j}^N \frac{x-x_i}{x_j-x_i}\,, \ i=0\,, ...\,, N\,.
\end{eqnarray}
	
Another crucial aspect of the PS approach is to operate in the Eddington-Finkelstein coordinates. In this coordinate system, Eq.\eqref{Sch_like_eq} linearizes with respect to the frequency $\omega$. In particular, working in the Eddington-Finkelstein coordinates offers enhanced simplicity in enforcing the boundary conditions.
In order to accomplish this objective, we implement the below transformations:
\begin{eqnarray}
	r\to \frac{r_h}{u} \ \ \text{and} \ \   \Psi=e^{i \omega r_*(u)}\psi\,.
\end{eqnarray}
Then, the wave equation \eqref{Sch_like_eq} transforms into the subsequent expression: 
\begin{eqnarray}\label{eq_EF}
	\psi''(u)+\left [\frac{f'(u)}{f(u)}+\frac{2ir_h\omega}{u^2f(u)}\right ]\psi'(u)-\left [\frac{l(l+1)}{u^2f(u)}+\frac{2ir_h \omega}{u^3f(u)}\right ]\psi(u)=0\,.
\end{eqnarray} 
It is evident that Eq.\eqref{eq_EF} satisfies the ingoing boundary condition at the horizon. Moreover, we implement the subsequent transformation to ensure that the Eq.\eqref{eq_EF} adheres to the outgoing boundary condition at the boundary:
\begin{eqnarray}
	\psi(u)=e^{\frac{2ir_h\omega}{u}}u^{2ir_h\omega f'(0)}\delta \psi(u)\,.
\end{eqnarray} 
Finally, we possess the subsequent scalar perturbation equation that fulfills the requirement of an ingoing boundary condition at the horizon and an outgoing boundary condition at the boundary:
\begin{eqnarray}\label{QNMs_eq}
	\delta\psi''(u)+\lambda_0(u)\delta\psi'(u)+s_0(u)\delta\psi(u)=0\,,
\end{eqnarray}
with
\begin{eqnarray}
	\lambda_0(u)&&=\frac{f'(u)}{f(u)}+\frac{4ir_h\omega f'(0)}{u}+\frac{2ir_h\omega}{u^2f(u)}-\frac{4ir_h\omega}{u^2}\,, \nonumber \\
	s_0(u)&&=-\frac{l(l+1)}{u^2f(u)}-\frac{2ir_h\omega[1+f(u)(-2+u f'(0))+uf'(u)(1-uf'(0))]}{u^3f(u)}\nonumber \\
	&&-\frac{4r_h^2\omega^2(-1+uf'(0))[1+f(u)(-1+uf'(0))]}{u^4f(u)}\,.
\end{eqnarray}
Afterward, the generalized eigenvalue equation can then be derived as follows:
\begin{eqnarray}\label{eq1}
	(M_0+\omega M_1)\psi=0,
\end{eqnarray}
where $M_i$ ($i=0,1$) denotes the linear combination of the derivative matrices. By directly solving the eigenvalue function, the QNFs can be ascertained.

	\bibliographystyle{style1}
	\bibliography{Ref}
\end{document}